\documentclass{ws-p8-50x6-00}

\usepackage{times}
\usepackage{amsmath}
\usepackage{amssymb}
\usepackage{mathptmx}

\newcommand{\um}[1]{\ensuremath{\,\mathrm{#1}}} 
\newcommand{\particle}[1]{\ensuremath{\, \mathrm{#1}}}
\newcommand{\antipart}[1]{\ensuremath{\, \overline{\mathrm{#1}}}}

\begin{document}

\title{The AMS-01 Time of Flight System}

\author{L.~Baldini, L.~Brocco, D.~Casadei, F.~Cindolo, A.~Contin,      \\
        G.~Laurenti, G.~Levi, A.~Margotti, A.~Montanari, F.~Palmonari, \\
        L.~Patuelli, C.~Sbarra, A.~Zichichi}

\address{Dipartimento di Fisica, Universit\`a di Bologna, via Irnerio
46, I-40126 Bologna, Italy,\\
and\\
INFN, Sezione di Bologna, viale Berti
Pichat 6/2, I-40127 Bologna, Italy}

\author{G. Castellini}

\address{CNR-IROE, Via Panciatichi 64, I-50127 Firenze, Italy}

\maketitle

\abstracts{The Time-of-Flight (TOF) system of the AMS detector gives
the fast trigger to the read out electronics and measures velocity,
direction and charge of the crossing particles.  The first version of
the detector (called AMS-01) has flown in 1998 aboard of the shuttle
Discovery for a 10 days test mission, and collected about $10^8$
events.  The new version (called AMS-02) will be installed on the
International Space Station on March 2004 and will operate for at
least three years, collecting roughly $10^{10}$ Cosmic Rays (CR)
particles.
The TOF system of AMS-01 successfully operated during the test
mission, obtaining a trigger efficiency better than 99.9\% and a time
resolution of 120 ps for protons and better for other CR ions.  In
addition, the TOF system was able to separate protons from all the
other CR nuclei within 1\% and to distinguish between downward and
upward crossing particles within at most $10^{-8}$.}

\section{Introduction}

The \emph{Alpha Magnetic Spectrometer} (AMS)~\cite{amsfirst} is a
particle detector that will be installed on the International Space
Station in 2004 to measure cosmic ray fluxes for at least three
years.

\begin{figure}[t]\centering
\includegraphics[width=\columnwidth]{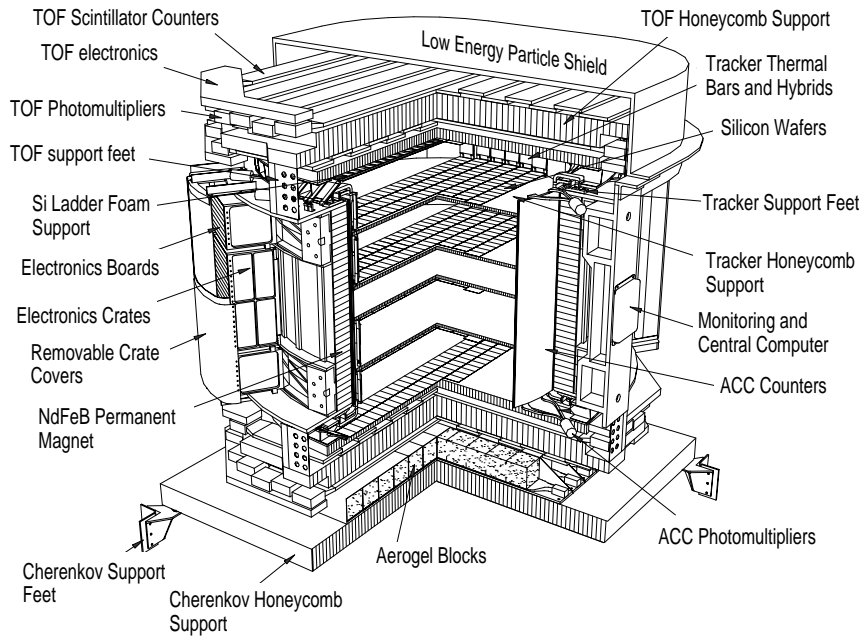}
\caption{The AMS detector for the STS-91 mission (AMS-1).}\label{AMS}
\end{figure}

During the precursor flight aboard of the shuttle Discovery (NASA
STS-91 mission, 2--12 June 1998), AMS collected data for about 180
hours.~\cite{amsall}  Figure~\ref{AMS} shows the detector (called
AMS-1 in the following), consisting of a permanent Nd-Fe-B magnet, six
silicon tracker planes, an anticoincidence scintillator counter
system, the time of flight (TOF) system consisting in four layers of
scintillator counters and a threshold aerogel \v{C}erenkov detector.

The TOF system~\cite{tof1} was completely designed and built at the
INFN Laboratories in Bologna. Its main goals are to provide the fast
trigger to AMS readout electronics, and to measure the particle
velocity ($\beta$), direction, position and charge.  In addition, it
had to operate in space with severe limits for weight and power
consuption.

Each TOF plane consists of 14 scintillator counters 1 cm thick
covering a roughly circular area of 1.6\um{m^2}.  The scintillation
light is guided to 3 Hamamatsu R5900 photomultipliers per side, whose
signals are summed to have a good redundancy and light collection
efficiency.  The total power consumption of the system (112 channels,
336 phototubes) was 150\um{W}, while its weight (support structure
included) was 250\um{kg}.

\section{The AMS-1 trigger}
The AMS-1 trigger logic consists of three levels.  The \emph{fast
trigger} (FT) processes the analog scintillators data and provides, in
about 50 ns, the zero time for the time-of-flight measurement. The
\emph{first level trigger} rejects events with hits on the
anticoincidence counter system and enhances the fraction of particles
crossing the tracker planes through the analysis of the pattern of hit
counters in the first and fourth TOF plane.  Finally, at the last
level the trigger logics suppress spurious fast triggers and finds
preliminary tracks on the silicon tracker by using the digitized data.

\begin{figure}[t]\centering
\includegraphics[width=0.7\columnwidth]{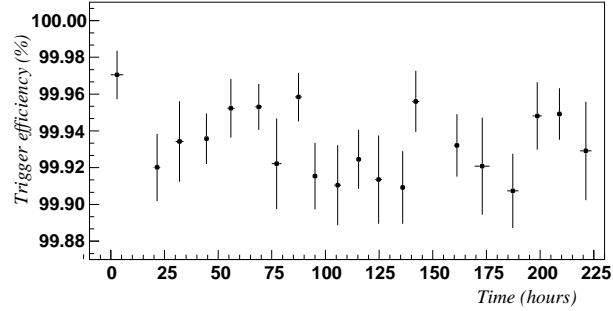}
\caption{Fast trigger efficiency during the STS-91 shuttle
flight.}\label{trig-eff-time}
\end{figure}

The FT signal is generated when at least one counter side in each TOF
plane produces a signal above a threshold corresponding to 40\% of a
minimum ionizing particle.  The efficiency of this selection criterion
could be measured with the same data taken during the STS-91 mission,
exploiting the characteristics of the TOF electronics, sensitive to
all particles impinging on the detector in an interval of about 16
$\mu$s around the trigger signal.  Up to eight hits can be registered
by each channel with a time resolution of 1 ns and a full charge
measurement.

The analysis of these unbiased data provided the instantaneous rate of
particles, the dead time and accidental rate, in addition to the total
FT efficiency.  Figure~\ref{trig-eff-time} shows that this efficiency,
for the whole duration of the STS-91 flight (data taken in the South
Atlantic Anomaly are excluded), was always above 99.9\%.

The background can be estimated by checking the consistency of the TOF
data and the trigger mask, and comes out to be about 0.5\% of the fast
triggers (due to electronics noise).  This background is completely
eliminated in the last level trigger by requiring the coincidence of
both sides of the same counter.

\section{Time of flight resolution}

The single channel time resolution is:~\cite{tof1}
\begin{equation}
   \sigma(x) = \sqrt{ \frac{\sigma_1^2}{N} + \frac{\sigma_2^2 x^2}{N}
		+ \sigma_3^2 } \; ,
\end{equation}
where $x$ is the distance of the particle crossing point from the
photomultiplier (PM), $N$ is the number of photons which convert on
the PM window, $\sigma_1$ depends upon the PM signal shape and the
trigger electronics, $\sigma_2$ takes into account the dispersion in
the photons path lengths and the constant term $\sigma_3$ depends on the
electronic noise at the low threshold discriminator input and on the
reference time dispersion on each channel.

The overall time resolution of a plane can be determined by measuring
the time of flight of ultrarelativistic particles between two given
planes, after correcting for the track length.

\begin{figure}[t]\centering
\begin{minipage}{0.52\textwidth}
\includegraphics[width=\columnwidth]{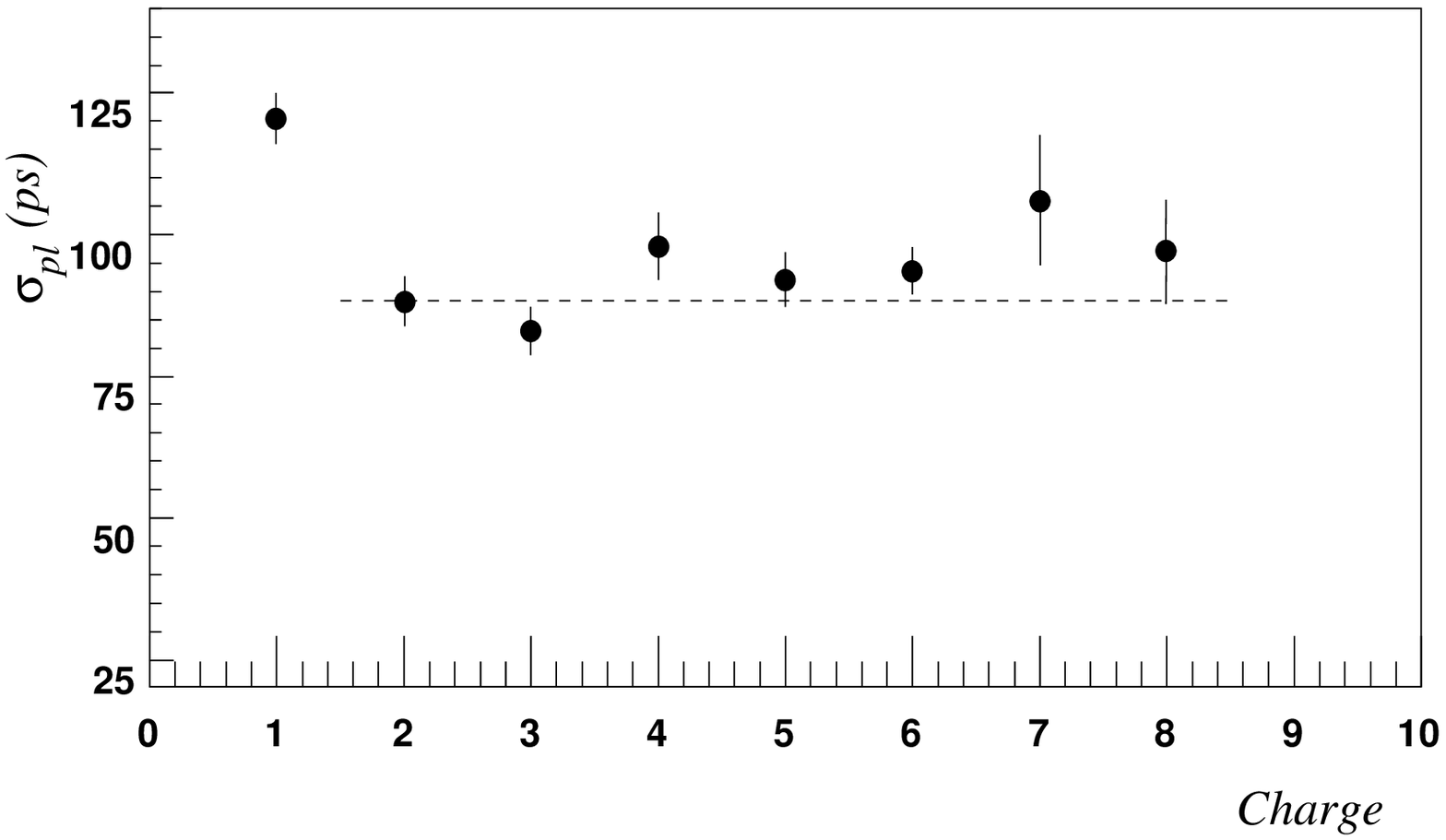}
\end{minipage}%
\begin{minipage}{0.48\textwidth}
\includegraphics[width=\columnwidth]{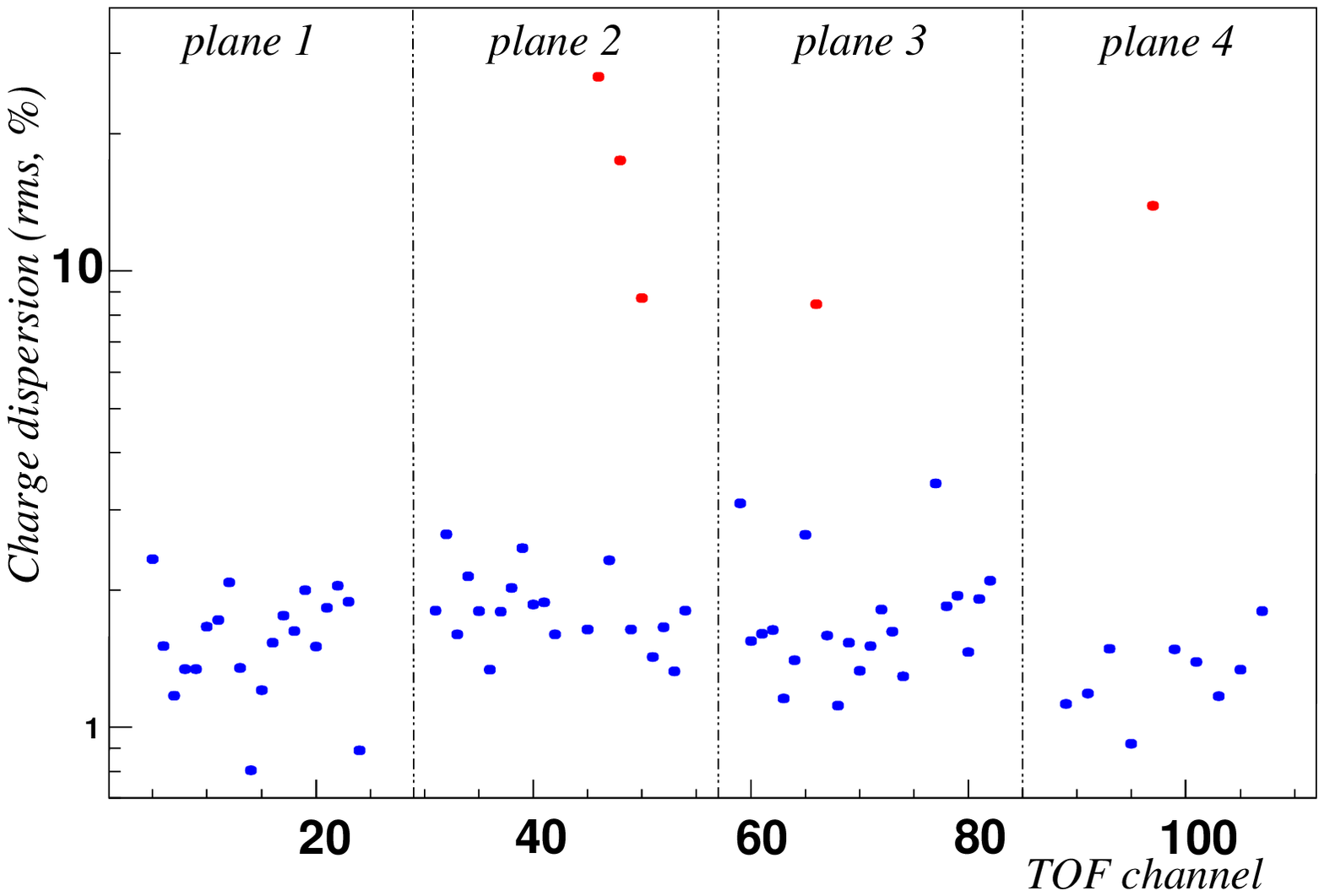}
\end{minipage}
\caption{\emph{Left}: single plane time resolution from the time of
flight between the first and the second ($\sigma_{12}$) or the third
($\sigma_{13}$) TOF plane, and the mean time resolution.
\emph{Right}: TOF charge peaks dispersion during the STS-91
flight.}\label{timeres}
\end{figure}

The time dispersion is expected to decrease with the nuclear charge
$Z$, due to the large number of photo-electrons produced by nuclei
with high atomic number, until it reaches the minimum value
$\sigma_3$.  Figure~\ref{timeres} (left panel) shows the single plane
time resolution as function of the particle charge: the horizontal
lines show that the limiting level $\sigma_3$ is 88\um{ps}.

\section{Photomultipliers stability}

The TOF system provides a measurement of the absolute charge of the
crossing particle in addition to the tracker, even if, due to the
strong constraints about power consuption, the TOF front-end
electronics was not optimized for energy deposition measurements.

The charge measurement was realized through a ``time-over-threshold''
method, whose response is proportional to the logarithm of the
deposited charge.  This method results in a good separating power
($\approx 5 \times 10^{-3}$) between singly and doubly charged
particles but has a poor charge resolution for $|Z|>2$.

The stability of the charge measurement was very good for all the 112
TOF channels, but five channels, as shown in figure~\ref{timeres},
right panel.

\section{Particle separation}

At the trigger level, one goal of the TOF system was to provide a
special flag for ions.  Accordingly, it was designed to distinguish in
a fast and efficient way cosmic ray protons from other nuclei.

One of the main purpose of the TOF system is the measurement of the
time of flight of the particles traversing the detector with a
resolution sufficient to distinguish upward from downward going
particles: an ``upward-going'' Helium nucleus wrongly labelled
``downward-going'' would be interpreted as an ``downward-going''
anti-Helium nucleus.

The average time of flight of the particles which traverse AMS is of
the order of 5 ns, while the time measurement has a resolution
$\sigma_t \lesssim 120$ ps, independent from the rigidity.  Thus
the probability to mistake the particle direction is well below
$10^{-11}$, the level needed for successful operation aboard the ISS,
where AMS is expected to collect at least $10^{10}$ events.

In addition, the velocity resolution of the TOF system, $\sigma
(\beta)/\beta\approx 3\%$, allows to discriminate
\particle{p}/\particle{e^+} and \antipart{p}/\particle{e^-} up to a
rigidity of 1.5 GV.

\end{document}